\documentclass[conference]{IEEEtran}
\IEEEoverridecommandlockouts
\usepackage{cite}
\usepackage{amsmath,amssymb,amsfonts}
\usepackage{algorithmic}
\usepackage{graphicx}
\usepackage{textcomp}
\usepackage{xcolor}
\usepackage{booktabs}
\usepackage{multirow}
\usepackage{tabularx}
\usepackage{subcaption}
\usepackage[table,xcdraw]{xcolor}

\def\BibTeX{{\rm B\kern-.05em{\sc i\kern-.025em b}\kern-.08em
    T\kern-.1667em\lower.7ex\hbox{E}\kern-.125emX}}
\begin{document}

\title{Finetuning LLMs for Automatic Form Interaction on Web-Browser in Selenium Testing Framework
}



\author{\IEEEauthorblockN{
    Nguyen-Khang Le\textsuperscript{1}\thanks{\textsuperscript{*} Corresponding author: lnkhang@jaist.ac.jp.}, 
    Hiep Nguyen\textsuperscript{1},
    Minh Ngoc Nguyen\textsuperscript{1},
    Son T. Luu\textsuperscript{1,3},
    Trung Vo\textsuperscript{1},\\
    Quan Minh Bui\textsuperscript{2},
    Shoshin Nomura\textsuperscript{2},
    Le-Minh Nguyen\textsuperscript{1}}

    \IEEEauthorblockA{
\textit{\textsuperscript{1}Japan Advanced Institute of Science and Technology, Ishikawa, Japan}\\
\textit{\textsuperscript{2}Amifiable Inc., Tokyo, Japan}\\
    }
    \textit{\textsuperscript{3}VNU University of Information Technology, Ho Chi Minh City, Vietnam}
}


\maketitle

\begin{abstract}
Automated web application testing is a critical component of modern software development, with frameworks like Selenium widely adopted for validating functionality through browser automation. Among the essential aspects of such testing is the ability to interact with and validate web forms, a task that requires syntactically correct, executable scripts with high coverage of input fields. Despite its importance, this task remains underexplored in the context of large language models (LLMs), and no public benchmark or dataset exists to evaluate LLMs on form interaction generation systematically. This paper introduces a novel method for training LLMs to generate high-quality test cases in Selenium, specifically targeting form interaction testing. We curate both synthetic and human-annotated datasets for training and evaluation, covering diverse real-world forms and testing scenarios. We define clear metrics for syntax correctness, script executability, and input field coverage. Our empirical study demonstrates that our approach significantly outperforms strong baselines, including GPT-4o and other popular LLMs, across all evaluation metrics. Our work lays the groundwork for future research on LLM-based web testing and provides resources to support ongoing progress in this area.
\end{abstract}


\section{Introduction}

Web applications are at the core of modern digital infrastructure, enabling critical services across domains such as finance, healthcare, education, and e-commerce. Ensuring their correctness and robustness is essential, as even small defects in user-facing components—especially web forms—can lead to serious usability issues or functional failures. Among the wide range of web functionalities, form interaction is particularly important and ubiquitous, facilitating tasks such as user registration, login, checkout, and data submission.

While frameworks like Selenium have become standard tools for automating web testing, they typically rely on manually authored test scripts. Writing these scripts is labor-intensive and error-prone, especially when dealing with modern, dynamic web forms that include conditional inputs, client-side validation, and real-time feedback. Despite the critical role of form testing, there exists no public dataset or benchmark that focuses on this specific task, limiting the development and evaluation of automated solutions. Moreover, existing large language models (LLMs), although capable of generating code, often struggle with producing syntactically correct, executable, and semantically accurate test cases for real-world web forms.

The emergence of large language models (LLMs), such as ChatGPT and GPT-4~\cite{openai2023gpt4}, has enabled new capabilities in automating complex tasks across domains. In web-based environments, LLM-powered agents have been developed to perform operations such as navigation and interaction by leveraging instruction-following and tool-use capabilities~\cite{autogpt2022, qin2023toollm, schick2023toolformer}. Several studies have investigated LLMs in web automation settings, introducing text-based browsing interfaces and strategies for managing verbose HTML structures through simplification and restructuring~\cite{nakano2021webgpt, zhou2023webarena, gur2023realworld, deng2023mind2web}. WebVoyager~\cite{he-etal-2024-webvoyager} further advances this line of work by incorporating multimodal LLMs into end-to-end automated web testing pipelines.

Despite these advances, most existing approaches primarily target high-level navigation or general-purpose web exploration. In contrast, fine-grained form interaction—such as accurately filling out input fields and generating executable scripts for testing frameworks like Selenium—remains underexplored. Existing systems often lack robustness in generating test scripts with comprehensive input coverage and execution reliability. Addressing this gap is essential for advancing LLM-driven web testing toward practical integration into software development pipelines.

In this paper, we introduce a novel approach for training large language models (LLMs) to generate high-quality form interaction test cases in Selenium. Our method emphasizes three key criteria: syntax correctness, executability, and input field coverage. To support this goal, we construct and release the first form interaction testing dataset, which includes both human-written and LLM-generated examples covering diverse real-world web forms. The contributions of this paper are as follows.
\begin{itemize}
    \item We propose a training approach for LLMs to generate executable Selenium test scripts that ensure syntax correctness and achieve high input field coverage.
    \item We develop a dataset for form interaction testing, combining human-annotated and synthetic web data.
    \item Extensive experiments demonstrate that our approach significantly outperforms strong baselines, including GPT-4o, across multiple evaluation metrics.
\end{itemize}

\begin{figure*}[ht]
    \centering
    \includegraphics[width=1\linewidth]{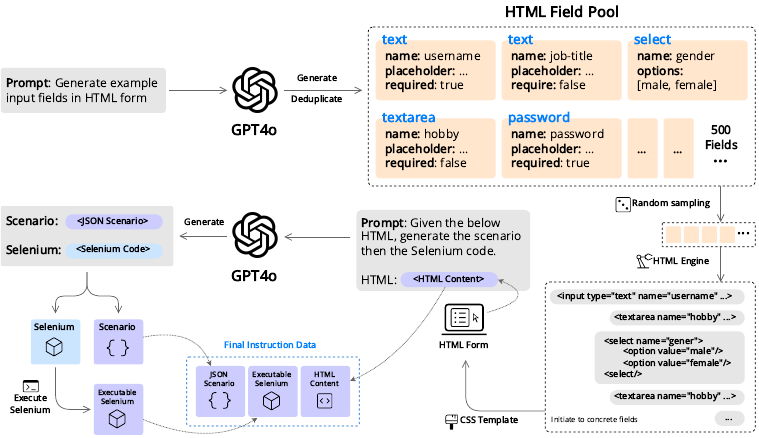}
    \caption{Overview of the data creation process.}
    \label{fig:llm4webtest}
\end{figure*}

\section{Related Work}

Web application testing is a fundamental yet resource-intensive task in modern software engineering. Traditional research has focused on both static and dynamic analysis techniques to ensure the functionality and security of web systems. Early efforts, such as \cite{ricca2001} and \cite{andrews2005}, laid the groundwork by exploring systematic approaches for improving test coverage and detecting faults through formal analysis and test case generation.

With the growing complexity of web applications, automation frameworks like Selenium have become essential tools for simulating user interactions and validating UI behavior. Despite their effectiveness, these tools typically rely on hand-crafted test scripts, which are time-consuming to author and maintain. Studies such as \cite{marchetto2012} have surveyed the inherent challenges in automating web application testing, emphasizing the need for scalable and low-maintenance solutions. More recent frameworks attempt to alleviate manual effort by recording user actions and translating them into test cases automatically \cite{smith2022}. While such techniques reduce scripting overhead, they still depend on access to human demonstrations and lack adaptability to unseen page structures or form behaviors.

In parallel, researchers have begun exploring the use of large language models (LLMs) to automatically generate test scripts. For instance, \cite{lee2022} proposed using generative models that adapt to changes in the application under test, enabling more resilient and dynamic testing workflows. WebVoyager \cite{he-etal-2024-webvoyager} further advances this direction by deploying multimodal LLM agents to explore and interact with live websites end-to-end. However, existing LLM-based methods often struggle with key aspects of form interaction, including syntax correctness of the generated scripts, execution reliability, and comprehensive input field coverage—especially in forms with conditional logic or validation constraints.

Our work addresses these challenges by introducing a dataset-driven approach to training LLMs for form interaction testing using Selenium. Unlike prior methods that treat form interaction as an auxiliary task, we treat it as a central objective, developing human-annotated and LLM-generated datasets specifically designed for evaluating and training models on script generation quality. By measuring and optimizing for syntax correctness, script executability, and input field coverage, our method achieves robust test case generation across diverse web forms. This approach complements and extends existing research by focusing on the intersection of LLMs and practical web testing needs—particularly in the domain of automated, reliable form filling. 



\begin{table*}
    \caption{Prompts used in GPT4o to generate the scenario and Selenium code.}
    \label{tab:prompt_used}
    \begin{tabularx}{\linewidth}{X} 
        \toprule
    
        Generate a test case scenario for filling in a given HTML form, providing step-by-step instructions, dummy data, the expected outcome. The test case scenario must be JSON-formatted.
    
    The goal is to ensure that each form element is filled appropriately, actions are conducted in the correct order.
    \# Steps
    
    1. **Form Analysis**: Break down the HTML form into its major fields (e.g., text inputs, radio buttons, dropdowns) and requirements. \\
    2. **Generate JSON Test Case**:
       - Identify each field by its name or identifier. \\
       - Identify if the field is required and only include required fields in the test case. \\
       - Include the html snippet for each form field. \\
       - Include appropriate dummy data for each form field and make sure the date is valid and follows the field's requirements. For entering date, use format 'mm/dd/yyyy'. \\
       - Specify the instructions for filling out each field in sequential order. \\
       - Mention the expected outcome (e.g., successful submission, error messages indicating missing required fields). \\
       - Exclude the submit button and use the `form.submit()` method for submitting forms instead of click button. \\
    
    \# Output Format
    
    **Test Case JSON**: 
       - Formatted in a structured JSON that describes the form fields, the data to be filled, step-by-step instructions, and the expected outcome.
    
    \#\#\# JSON Structure: [Json structure] \\
        \bottomrule
    \end{tabularx}

\end{table*}

\begin{figure*}[ht]
    \centering
    \begin{subfigure}[b]{0.4\linewidth}
        \includegraphics[width=\linewidth]{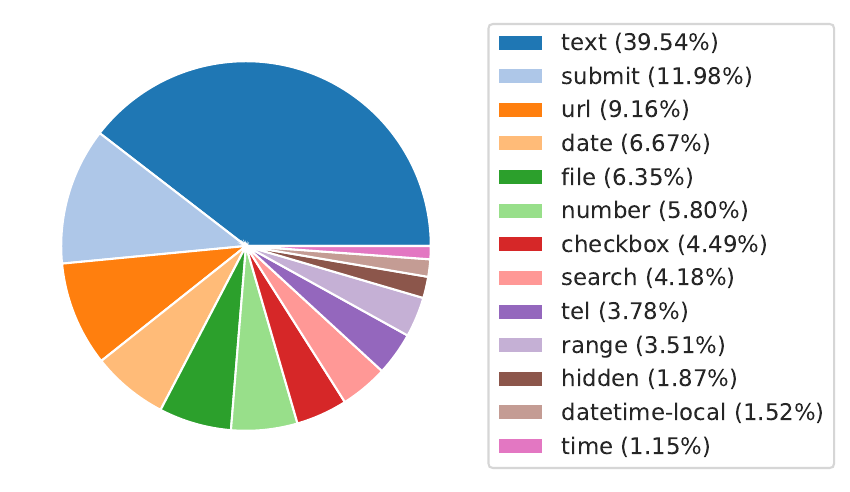}
        \caption{Distribution of field types.}
    \end{subfigure}
    \hfill
    \begin{subfigure}[b]{0.29\linewidth}
        \includegraphics[width=\linewidth]{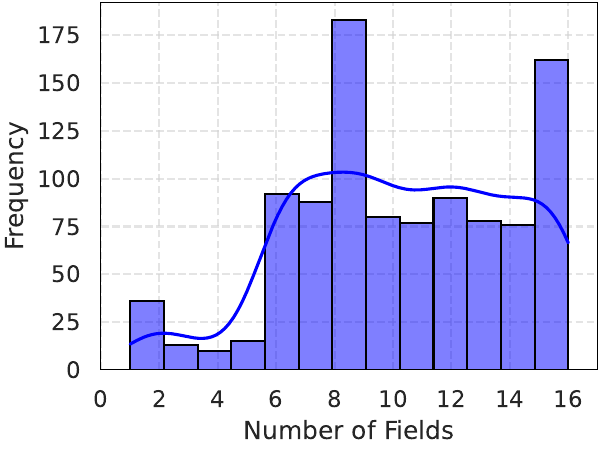}
        \caption{Distribution of number of fields.}
    \end{subfigure}
    \hfill
    \begin{subfigure}[b]{0.29\linewidth}
        \includegraphics[width=\linewidth]{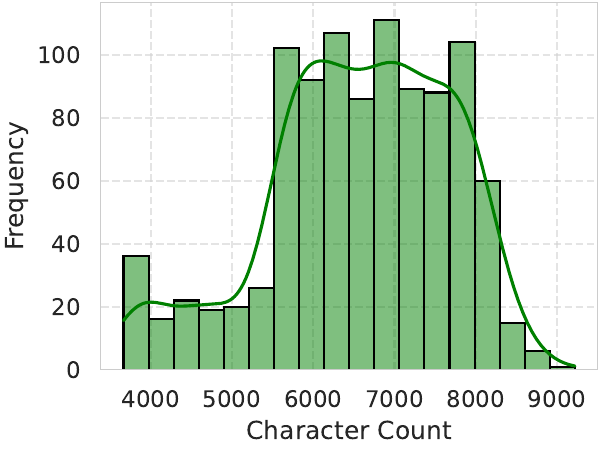}
        \caption{Distribution of number of characters.}
    \end{subfigure}
    \caption{Statistics of the synthetic HTML in training data.}
    \label{fig:html_statistics}
\end{figure*}

\section{Methodology}
\subsection{Training Data Creation}

To generate high-quality instruction data for training a large language model (LLM) to produce test scenarios and corresponding Selenium code, we design a multi-stage pipeline centered around GPT-4o, as illustrated in Figure~\ref{fig:llm4webtest}.

The process begins by prompting GPT-4o to generate diverse input fields commonly found in HTML forms, such as \texttt{<input>}, \texttt{<textarea>}, \texttt{<select>}, and \texttt{<password>}. Each field is annotated with attributes such as \texttt{name}, \texttt{placeholder}, \texttt{required}, and, in the case of selection fields, a list of \texttt{options}. Redundant entries are removed to create a clean and diverse HTML Field Pool containing 500 unique form field configurations.

A random sampling strategy is then applied to select a subset of fields from this pool. These fields are assembled into a concrete HTML form using an HTML engine and styled with a predefined CSS template to enhance realism. Figure \ref{fig:html_statistics} shows the statistics of the generated HTML. This generated HTML form serves as the input to a second GPT-4o prompt. 

\subsection{Test Scenario and Selenium Generation}

We prompt GPT-4o to simultaneously generate both the test scenario and the corresponding Selenium code from a given HTML form. Table \ref{tab:prompt_used} shows the prompts used in GTP-4o. GPT-4o then generates two outputs: (1) a natural language scenario represented in JSON format and (2) the corresponding Selenium code that automates interaction with the form.  The scenario is structured in JSON format and includes a detailed description of the user action, the relevant input fields, their identifiers (e.g., by \texttt{id}, by \texttt{name}), the interaction method, the instruction, and the expected outcome. Based on this structured scenario, GPT-4o proceeds to generate the Selenium script. The generated code is then executed using a web engine; scripts that fail to execute are discarded, while only executable ones are retained for further use.

The final instruction data consists of three components: the HTML content, the scenario in JSON, and the executable Selenium code. This structured and realistic dataset enables the LLM to learn to generate both test scenarios and automation scripts grounded in actual web content. The key advantage of this process lies in its ability to synthesize high-quality, executable, and semantically consistent data by leveraging LLMs and controlled sampling mechanisms.


\section{Experiments}

We evaluate our approach on the task of generating Selenium scripts for form interaction testing. Our experiments assess the ability of large language models (LLMs) to produce scripts that are syntactically correct, executable, and achieve high coverage of input fields. Below, we describe the models, datasets, and evaluation metrics used in our study.

\subsection{Models}

\textbf{GPT Family:} We compare our approach with the GPT family models, including \textit{GPT4o} and \textit{GPT-OSS} (20B and 120B versions), which serve as a strong baseline for code generation.

\textbf{State-of-the-art LLMs}: We use the families of \textit{Qwen2.5}\cite{bai_qwen_2023}, \textit{Qwen3}\cite{yang2025qwen3technicalreport}, and \textit{Llama3.1}\cite{grattafiori2024llama3herdmodels}, which are the recent state-of-the-art open-source language models with strong performance on code generation tasks. We employ their instruction-tuned variants and fine-tune them on our synthetic data of form interaction. We compare the performance of the original model with the model finetuned using our approach. 

\textbf{Code generation LLMs}: We use the model \textit{Qwen2.5 Coder Instruct}\cite{hui2024qwen25codertechnicalreport}. We fine-tune them on our synthetic data of form interaction. We compare the performance of the original model with the model finetuned using our approach.

\definecolor{mygray}{RGB}{230, 230, 230}
\definecolor{myblue}{RGB}{204, 229, 255}
\definecolor{myred}{HTML}{FFEBEE}

\begin{table*}[ht!]
\caption{Performance comparison on real-world form interaction testing. \textsc{Our approach} is evaluated under two settings: with filtered training data that excludes unexecutable Selenium code; and without filtering (w/o filter), where all Selenium code is used for training.
}
\label{tab:main_result}
\centering
    \begin{tabular}{llccc}
\toprule
\textbf{Model} & \textbf{Run} & \textbf{Syntax Correctness (\%)} & \textbf{Executability (\%)} & \textbf{Input Coverage (\%)} \\
\midrule

\multicolumn{5}{c}{\textit{GPT Family (Baseline)}} \vspace{1mm}\\ 
\rowcolor{mygray}
GPT-4o & Original & 97.67 & 50.38 & 39.28 \\
\rowcolor{mygray}
GPT-OSS (120B) & Original & 100 & 45.74 & 25.07 \\
\rowcolor{mygray}
GPT-OSS (20B) & Original & 91.47 & 46.51 & 23.74 \\
\midrule

\multicolumn{5}{c}{\textit{Qwen2.5 Family of models}} \vspace{2mm}\\

\multirow{3}{*}{Qwen2.5-14B (Instruct)} 
& Original & 99.22 & 49.61 & 39.35 \\
& \textsc{Our approach} (w/o filter) & \textbf{100.00} &58.91 & 39.79 \\
\rowcolor{myblue}
& \textsc{Our approach} & 99.22 & \textbf{59.69} & \textbf{40.53} \vspace{3mm}\\

\multirow{3}{*}{Qwen2.5-7B (Instruct)} 
& Original & 99.22 & 48.84 & 4.96 \\
& \textsc{Our approach} (w/o filter) & 59.38 & 23.26 & 16.26 \\
\rowcolor{myblue}
& \textsc{Our approach}  & 100.00 & 48.84 & 23.68 \\
\midrule

\multicolumn{5}{c}{\textit{Qwen3 Family of models}} \vspace{2mm}\\ 
\multirow{3}{*}{Qwen3-14B (Instruct)} 
& Original & 97.67 & 48.06 & \textbf{37.83} \\
& \textsc{Our approach} (w/o filter) & 98.44 & 60.47 & 33.15 \\
\rowcolor{myblue}
& \textsc{Our approach} & \textbf{100.00} & \textbf{60.47} & 35.96 \vspace{3mm}\\

\multirow{3}{*}{Qwen3-8B (Instruct)} 
& Original & 100.00 & 17.24 & 19.12 \\
& \textsc{Our approach} (w/o filter) & 98.41 & 28.57 & 27.38 \\
\rowcolor{myblue}
& \textsc{Our approach} & \textbf{100.00} & 25.20 & 21.44\\
\midrule

\multicolumn{5}{c}{\textit{Llama Family of models}} \vspace{2mm}\\ 
\multirow{3}{*}{Llama3.1-8B (Instruct)} 
& Original & 94.57 & 34.11 & 0.45 \\
& \textsc{Our approach} (w/o filter) & \textbf{99.22} & 39.53 & 34.24 \\
\rowcolor{myblue}
& \textsc{Our approach} & 98.45 & \textbf{45.74} & \textbf{36.80} \\

\midrule
\multicolumn{5}{c}{\textit{Code Generation Models}} \vspace{2mm}\\ 
\multirow{3}{*}{Qwen2.5-7B-Coder (Instruct)} 

& Original & 96.80& 56.80 & 10.89 \\
& \textsc{Our approach} (w/o filter) & \textbf{100.00}& 57.85 & \textbf{20.12} \\
\rowcolor{myblue}
& \textsc{Our approach}  & 98.31 & \textbf{66.10} & 15.40\\



\bottomrule

\end{tabular}
\end{table*}

\subsection{Datasets}

We evaluate the models on two test sets:

\begin{itemize}
    \item \textbf{Synthetic Forms:} A curated collection of 500 HTML forms programmatically generated to cover a wide range of input types (e.g., \texttt{text}, \texttt{email}, \texttt{password}, \texttt{checkbox}, \texttt{radio}, \texttt{select}), layout patterns, and interaction complexity (e.g., required vs. optional fields, conditional logic).
    
    \item \textbf{Real-World Forms:} 133 web forms collected from publicly available websites, including login, registration, and contact forms. These forms reflect real-world HTML structure diversity and domain-specific constraints.
\end{itemize}


\subsection{Evaluation Metrics}

To measure the quality of generated scripts, we use the following three metrics:

\begin{itemize}
    \item \textbf{Syntax Correctness:} Whether the generated script is syntactically valid. We validate Selenium scripts using language-specific syntax checkers.

    \item \textbf{Executability:} Whether the script runs without errors in a headless browser environment (e.g., Chromium). This checks runtime correctness and functional validity.

    \item \textbf{Input Coverage:} The proportion of input fields in the form that are correctly identified and filled by the script. This metric captures the completeness of the test case in terms of form interaction.
\end{itemize}


\section{Results}

\begin{table}[t!]
    \centering
    \caption{Performance comparison on Synthetic Form interaction testing, measured on syntax correctness (Syntax), executability (Execute), and input coverage (Cover).}
    \begin{tabular}{llccc}
    \toprule
    \textbf{Model} & \textbf{Setting} & \textbf{Syntax} & \textbf{Execute} & \textbf{Cover} \\
    \midrule
    \textbf{GPT-4o} & API & 99.60 & 74.68 & 76.40 \\
    \midrule
    \multirow{3}{*}{\textbf{Qwen2.5-14B-Inst}} 
    & Original & 96.40 & 77.40 & 77.97 \\
    & \textsc{Our approach} & 100.00 & \textbf{82.00} & \textbf{86.38} \\
    \midrule
    \multirow{2}{*}{\textbf{Qwen3-14B}} 
    &  Original & 100.00 & 65.20 & 85.30 \\
    &  \textsc{Our approach} & 100.00 & \textbf{81.60} & \textbf{87.18} \\
    \bottomrule
    \end{tabular}
    \label{tab:simple_form_result}
\end{table}

\subsection{Main Results}

Table~\ref{tab:main_result} presents the performance comparison on real-world form interaction testing. Across different model families, our approach consistently improves executability and input coverage compared to the original baselines, while maintaining high syntax correctness. The filtering strategy proves effective in eliminating noisy or unexecutable samples, leading to more reliable script generation that runs successfully and covers a broader range of input fields. Importantly, the improvements are observed across diverse backbones, including Qwen, Llama, and code generation models, demonstrating that our method generalizes well beyond a single model family. These results confirm the effectiveness of our approach in enhancing LLM-based form interaction testing in realistic web environments.


Table~\ref{tab:simple_form_result} reports the performance on synthetic form interaction testing. Across all models, our approach consistently improves executability and input field coverage compared to the original settings, while maintaining perfect syntax correctness. This indicates that our approach leads to more reliable script generation that not only runs successfully but also achieves broader coverage of input fields. Notably, the improvements hold across different model backbones, demonstrating the robustness and generality of our approach in enhancing LLM-based form interaction testing.

\subsection{Analysis of Error rate in different field types}

\begin{figure}[ht]
    \centering
    
    \begin{subfigure}[b]{\linewidth}
        \includegraphics[width=\linewidth]{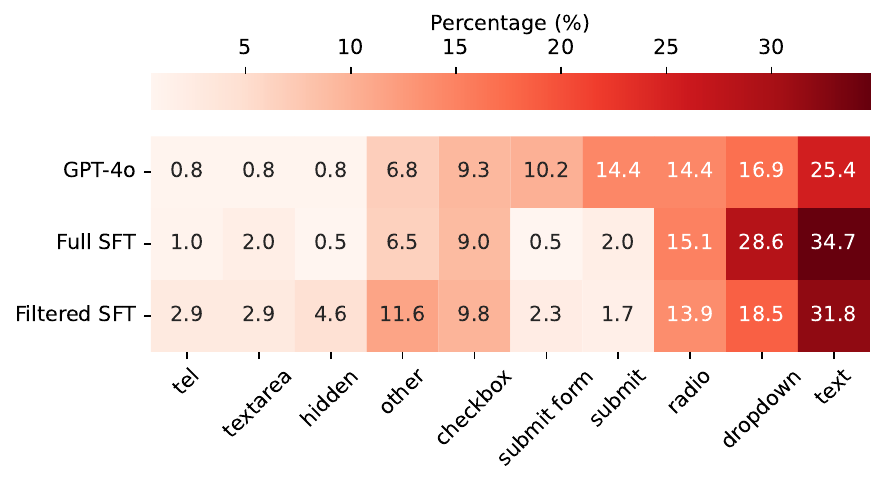}
        \caption{\textbf{Qwen2.5-14B-Instruct}'s Percentage of error in each field type.}
        \vspace{5mm}
        \label{fig:field_error_qwen14b}
    \end{subfigure}
    
    \begin{subfigure}[b]{\linewidth}
        \includegraphics[width=\linewidth]{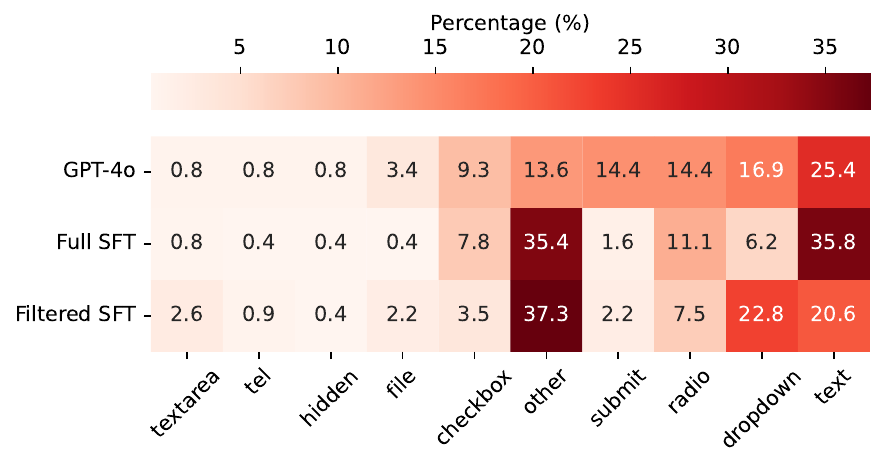}
        \caption{\textbf{Qwen2.5-7B-Instruct}'s Percentage of error in each field type.}
        \label{fig:field_error_qwen7b}
    \end{subfigure}

    \begin{subfigure}[b]{\linewidth}
        \includegraphics[width=\linewidth]{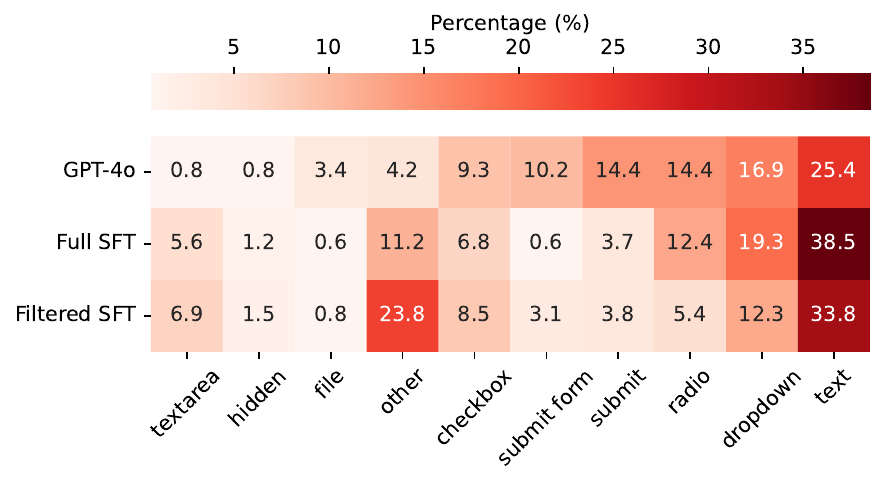}
        \caption{\textbf{Qwen3-14B-Instruct}'s Percentage of error in each field type.}
        \label{fig:field_error_qwen3_14b}
    \end{subfigure}
    \caption{Percentage of errors across field types for various LLMs. "Filtered SFT" refers to models fine-tuned with our filtering step, where unexecutable Selenium code is excluded, while "Full SFT" denotes fine-tuning without filtering, using all Selenium code for training.}
    \label{fig:field_error}
\end{figure}

Figure~\ref{fig:field_error} shows the percentage of form field errors across different HTML field types for the GPT-4o model, and baseline LLMs supervised fine-tuning with our approach (Filtered SFT), and fine-tuning without filtering, where all Selenium code is used (Full SFT). In Qwen2.5-14B-Instruct (Figure~\ref{fig:field_error_qwen14b}), all three models exhibit relatively high error rates in complex field types such as \texttt{text}, \texttt{dropdown}, and \texttt{radio}, while achieving low error rates on simpler or infrequently used fields like \texttt{tel}, \texttt{textarea}, and \texttt{hidden}. The Filtered SFT variant displays a more even distribution of errors, reducing failure in common fields such as \texttt{submit} and \texttt{radio} compared to the Full SFT baseline. This suggests that data filtering may help reduce overfitting to noisy or rare patterns. GPT-4o shows overall lower error percentages and maintains consistency across field types, serving as a reference point for generalization performance.

In the Qwen2.5-7B-Instruct (Figure~\ref{fig:field_error_qwen7b}) and Qwen3-14B-Instruct (Figure~\ref{fig:field_error_qwen3_14b}), similar trends emerge, but with sharper contrasts between field types. Notably, Full SFT and Filtered SFT both struggle significantly with the \texttt{other} and \texttt{text} categories, indicating that smaller models may be more sensitive to ambiguous field semantics. The Filtered SFT variant again shows improved robustness in fields such as \texttt{submit} and \texttt{radio}, although its error rate on \texttt{dropdown} slightly increases compared to the Full SFT model. Interestingly, GPT-4o's error pattern remains stable across both scales, reinforcing the observation that model size and fine-tuning strategy can introduce distinct biases in field-specific performance. Comparing the two model scales, Qwen2.5-14B exhibits more stable performance across field types, while Qwen2.5-7B shows more pronounced spikes in error for specific categories.

\subsection{Analysis of Causes of Errors}
Across both model scales and all training regimes, the dominant failure mode underlying the per–field-type error profiles in Figure~\ref{fig:field_error} is \emph{inconsistent identification of the intended target field} within the active interaction context (typically, the currently focused form). Concretely, the model often predicts a correct \emph{semantic intent} (e.g., ``fill email''), yet binds that intent to an \emph{incorrect DOM element}. This manifests as: (i) selecting a visually adjacent but non-interactable wrapper (e.g., a \texttt{<div>} around an \texttt{<input>}); (ii) selecting a hidden or disabled control; (iii) picking the wrong member of a group (\texttt{radio}/\texttt{checkbox}); (iv) confusing native \texttt{<select>} with a custom dropdown built from \texttt{<div>}/\texttt{<li>}; or (v) leaving the form scope and hitting a similarly named field elsewhere on the page. Typical Mis-Bind patterns include:
\begin{itemize}
    \item \textbf{Text inputs} (text, email, number): Choosing a placeholder -only field, a hidden clone used for masking, or a sibling used for validation.
    \item \textbf{Dropdowns}: Treating a custom combobox as a native \texttt{<select>} (or vice versa). 
    \item \textbf{Radio/Checkbox groups}: Clicking the first matching input by \texttt{@type} or \texttt{@name}, ignoring the label/value.
    \item \textbf{Submit}: Clicking a non-submit button (e.g., a reset or a tab switch) or a disabled button before client-side validation completes. 
    \item \textbf{Hidden/Offscreen}: Selecting the elements with attributes \texttt{display:none}, \texttt{visibility:hidden}, zero-size, or off-canvas.
    \item \textbf{Context leakage}: Leaving the current form and interacting with a homonymous field elsewhere (e.g., header search).
\end{itemize}

\subsection{Examples of Faulty Selenium Snippets and Implications}

To better illustrate how incorrect field identification translates into faulty automation scripts, we analyze common failure patterns observed in generated Selenium code. The following examples reflect real mis-bindings found in the evaluation set, highlighting the subtle but critical mismatches between semantic intent and DOM binding.

\begin{itemize}
    \item \textbf{Example 1: Misbinding to Wrapper instead of Input}
\begin{verbatim}
driver.find_element(By.XPATH, 
  "//div[@id='mail']").send_keys("a@a")
\end{verbatim}
    This line targets a \texttt{<div>} that visually wraps the input field (e.g., for styling or layout). Since \texttt{<div>} is not an interactable input element, this action raises a \texttt{ElementNotInteractableException}.  
    \item \textbf{Example 2: Selecting Disabled or Hidden Field}
\begin{verbatim}
driver.find_element(By.CSS_SELECTOR, 
  "input[name='tel']").send_keys("123")
\end{verbatim}
    The matched element exists in the DOM but is either \texttt{disabled} or hidden (e.g., \texttt{display:none}). No visible feedback or state change occurs on the page. 
    \item \textbf{Example 3: Context Leakage}
\begin{verbatim}
driver.find_element(By.NAME, 
  "search").send_keys("query")
\end{verbatim}
    If multiple fields share the same name (e.g., one in the form, one in the navbar), the global selector may match the wrong instance. 
\end{itemize}



\section{Conclusions}

This work presents a comprehensive framework for form-centric web automation using large language models, addressing a critical yet underexplored area in automated web testing. By introducing both a novel data generation pipeline and robust evaluation metrics, we enable systematic training and assessment of LLMs on form interaction tasks. Our curated datasets and empirical results demonstrate that LLMs can be effectively adapted to generate syntactically correct, executable, and high-coverage Selenium scripts. The significant performance improvements over strong baselines highlight the potential of LLMs in this domain. We hope our contributions—benchmarks, datasets, and methodology—serve as a foundation for further research into LLM-based software testing and broader applications in web automation.

\bibliographystyle{IEEEtran} 
\bibliography{references} 

\end{document}